# 3M: An Effective Multi-view, Multi-granularity, and Multi-aspect Modeling Approach to English Pronunciation Assessment

Fu-An Chao[1], Tien-Hong Lo[1,2], Tzu-I Wu[2], Yao-Ting Sung[3], Berlin Chen[2]

[1] Research Center for Psychological and Educational Testing, National Taiwan Normal University, Taiwan
[2] Department of Computer Science and Information Engineering, National Taiwan Normal University, Taiwan
[3] Department of Educational Psychology and Counseling, National Taiwan Normal University, Taiwan

{fuann, teinhonglo, 61047087s, sungtc, berlin}@ntnu.edu.tw

*Abstract*—As an indispensable ingredient of computer-assisted pronunciation training (CAPT), automatic pronunciation assessment (APA) plays a pivotal role in aiding self-directed language learners by providing multi-aspect and timely feedback. However, there are at least two potential obstacles that might hinder its performance for practical use. On one hand, most of the studies focus exclusively on leveraging segmental (phonetic)-level features such as goodness of pronunciation (GOP); this, however, may cause a discrepancy of feature granularity when performing suprasegmental (prosodic)-level pronunciation assessment. On the other hand, automatic pronunciation assessments still suffer from the lack of large-scale labeled speech data of non-native speakers, which inevitably limits the performance of pronunciation assessment. In this paper, we tackle these problems by integrating multiple prosodic and phonological features to provide a multi-view, multi-granularity, and multi-aspect (3M) pronunciation modeling. Specifically, we augment GOP with prosodic and self-supervised learning (SSL) features, and meanwhile develop a vowel/consonant positional embedding for a more phonology-aware automatic pronunciation assessment. A series of experiments conducted on the publicly-available speechocean762 dataset show that our approach can obtain significant improvements on several assessment granularities in comparison with previous work, especially on the assessment of speaking fluency and speech prosody.

## I. INTRODUCTION

To tie in with the trend of the rapid growth of globalization, self-directed foreign language learning systems have become more and more appealing. As one of alternative pathways, computer-assisted pronunciation training (CAPT) has emerged and become an effective aid for non-native (L2) speakers to learn different foreign spoken (L1) languages. In contrast to the conventional curriculum in school, CAPT is much more time-efficient and cost-effective that can provide immediate feedback to L2 learners on their pronunciation quality, so as to help promote their speaking proficiency. In addition, standardized test services also deploy CAPT systems to assist in the proficiency evaluation of L2 speakers, such as TOFEL [1], which have received broad success in bridging the gap between academic and commercial sectors.

In light of the potential value of CAPT, there have been exhaustive studies on the development of related techniques, which may fall roughly into two categories: mispronunciation detection and diagnosis (MDD) and automatic pronunciation assessment. In order to detect the pronunciation patterns of non-native language learners, the majority of efforts have been made on the scoring of phone-level pronunciation quality by leveraging techniques originated from automatic speech recognition (ASR) [2][3][4][5][6]. Among these techniques, goodness of pronunciation (GOP) based algorithm and its variants are the most celebrated instantiations, which have been empirically shown to correlate well with the human assessment [4][6].

Compared with MDD, the aim of automatic pronunciation assessment is to give feedback to L2 learners on some specific aspects or traits of their spoken language usage. Although precise detection of mispronounced phone- or word-segments is not essential [7], providing comprehensive feedback for L2 learners is more crucial in the pronunciation assessment, where different speech granularities and perspectives of phonological patterns must be considered. For granularity, there are only a dearth of literature exploring multi-granularity assessment [11][12], but most of the previous work only provided a single score for each granularity. On the other hand, while other pronunciation aspects such as fluency [8], lexical stress [9] and intonation [10] are well-studied, these features are often modeled independently. Recently, goodness of pronunciation Transformer (GOPT) was introduced in [13], which advocated a multi-aspect and multi-granularity processing regime for pronunciation assessment. Although GOPT has achieved promising results on speechocean762 (a publicly-available benchmark dataset curated for research on CAPT), GOPT still suffers from three deficiencies that may make it suboptimal for multi-aspect pronunciation assessment: First, the GOP-based features adopted by GOPT solely capture segmental-level pronunciation characteristics which focuses mainly on the pronunciation scores of the phonetic segments and neglects the contextual influence between actually uttered phones (e.g., liaison and omission), thereby resulting in a poor grasp of other suprasegmental-level pronunciation patterns (e.g., stress, fluency and prosody). Second, the canonical phone embedding employed in GOPT is merely one-hot encoding, which falls short in providing more representative phonological features that are anticipated to promote the assessment accuracy. Third, automatic pronunciation assessment of L2 learners is still faced with the resource-scarcity problem which will dramatically limit the model performance. As such, how to capitalize on meaningful and robust knowledge distilled from pre-trained models emerges as a prominent issue in automatic pronunciation assessment.

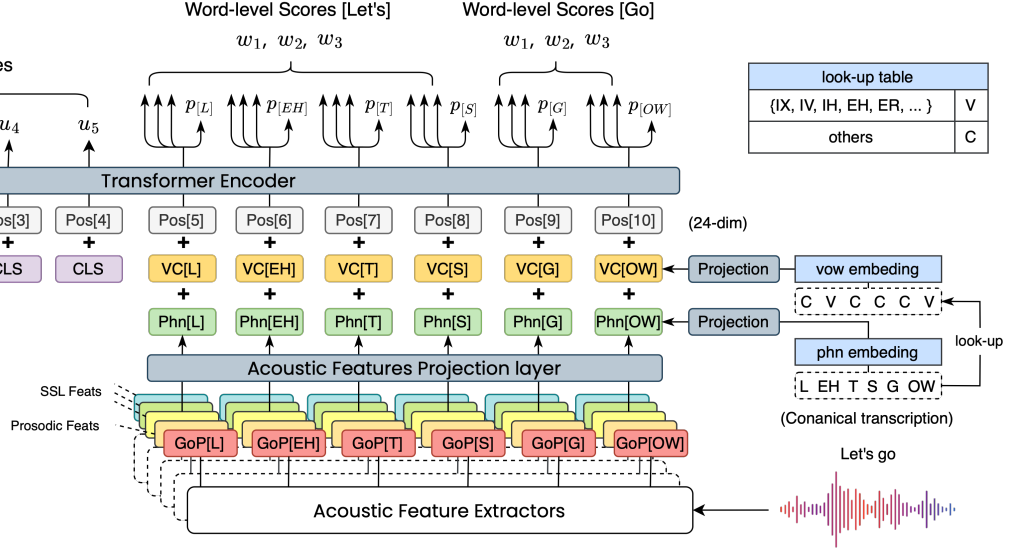

**Fig 1.** A schematic diagram of proposed 3M model for automatic pronunciation assessment,
$\{p_{[L]}, p_{[EH]}, p_{[T]}, p_{[G]}, p_{[OW]}\}$ refers to the phone-level scores of {L, EH, T, S, G, OW},
$\{w_1, w_2, w_3\}$ refers to word-level scores {accuracy, stress, total},
$\{u_1, u_2, u_3, u_4, u_5\}$ refers to utterance-level scores {accuracy, fluency, completeness, prosody, total}

In the context of the above background, we propose a unified modeling approach that considers multi-view, multi-aspect and multi-granularity for pronunciation assessment (dubbed the 3M model). The main contributions of the 3M model are at least three-fold: 1) we incorporate prosodic features apart from GOP into the GOPT framework, 2) a simple yet effective vowel/consonant positional embedding scheme is employed to help render phonological patterns in an effort to predict high-level pronunciation scores accurately, and 3) to work around the inherent under-resourced issue, we seek to leverage contextualized acoustic representations, (viz. self-supervised learning (SSL) embeddings) to enhance the 3M model. Extensive experiments on a benchmark English CAPT dataset demonstrate that our proposed approach can yield considerable improvements over GOPT on several assessment granularities, especially for scoring the speaking fluency and speech prosody of L2 learners.

The remainder of this paper is organized as follows. Section 2 recapitulates the GOPT model which acts as the archetype model of our method and elucidates the details of our proposed 3M model. Section 3 presents the experimental setup, results and associated discussion. In Section 4, we conclude the paper and envisage possible future research directions.

## II. METHODOLOGY

### A. GOPT

To the best of our knowledge, in contrast to conventional approaches which usually model different aspect of pronunciation scores independently, GOPT is among the seminal studies that jointly predict multi-aspect scores of pronunciation assessment with different granularities. The input GOP feature of GOPT is a combination of log phone posterior (LPP) [14] and log posterior ratio (LPR) [15] that can be defined as follows:

$$LPP(p) = \log p(p|\mathbf{o}; t_s, t_e)$$
$$\approx \frac{1}{t_e - t_s + 1} \sum_{t=t_s}^{t_e} \log p(p|o_t) \quad (1)$$

where $p$ is the canonical phone, $\mathbf{o}$ is input observations, and $t_s$ and $t_e$ are the start and end frame indexes, while the LPR value between $p_j$ and $p_i$ is defined as:

$$LPR(p_j, p_i) = \log p(p_j|\mathbf{o}; t_s, t_e) - \log p(p_i|\mathbf{o}; t_s, t_e) \quad (2)$$

Given an inventory of $M$ phones, the GOP feature with respect to the segment $i$ of a learner's utterance that corresponds to a canonical phone $p_i$ in a text prompt can be derived as follows:

$$[LPP(p_1)\ldots LPP(p_M), LPR(p_1|p_i)\ldots LPR(p_M|p_i)]^T \quad (3)$$

Instead of using a hierarchical model structure to obtain utterance-level representations, GOPT designs five trainable

[CLS] tokens which is spliced together with the phone-level input sequence, along a similar vein to the modeling paradigm in BERT [16], so as to obtain five aspects of utterance-level pronunciation scores. Through the context of the Transformer encoder [17], GOPT can produce utterance-level representations corresponding to each [CLS] token that has contextualized information among phones and between each phone and the whole utterance.

*B. 3M*

When it comes to multi-aspect pronunciation assessment, suprasegmental-level scoring and feedback information (e.g., pronunciation quality pertaining to fluency, prosody, stress, and others) are critical for the L2 learners (e.g., fluency, stress score, and others). However, the GOP-based features merely offer the segmental-level (viz. phone-level) information alone, which is insufficient for the complete modeling of multi-aspect pronunciation assessment. Hence, to provide multi-view information, we explore multiple prosodic and phonological features as auxiliary cues to facilitate the process of model learning for different aspects of (viz. suprasegmental-level) pronunciation assessment. In support of this objective, we extend the idea of GOPT by developing a novel neural model that is more amenable holistic pronunciation assessment, termed 3M. As shown in Figure 1, 3M is characterized by its multi-view, multi-aspect and multi-granularity modeling paradigm for pronunciation assessment. In the following, we describe the ingredient components of 3M.

- **Prosodic Features**

As for suprasegmental-level information, we investigate two types of prosodic features: duration and energy, both of which are essential indicators to determine the pronunciation quality of stress, fluency, and prosody.

***Duration***: conventional GOP features do not convey information about the duration of each phone- and word-level segment pronounced by an L2 learner, due to the average operation over all time frames, as shown in Eq (1). Nevertheless, the phone-level duration statistics often reflect the perception of a speaker's rhythm, which is highly correlated to the speaking proficiency of a learner. For instance, [18] found that duration is one of the most important features for the classification task on predicting the proficiency of French learners speaking German, which is also consistent with the linguistic literature [19][20]. As the result, we calculate the duration of each phone segment and take as an 1-dimensional feature to augment the original acoustic input.

***Energy***: energy is another important feature of speech detection; the energy distribution may be related to the intonation property. In out experiments, instead of using the absolute value of energy which has shown to have no strong correlation with the quality of pronunciation [21]. Alternatively, we derive the root-mean-square energy (RMSE) suggested by [22] and calculate the corresponding statistic as another prosodic feature for each phone segment. Specifically, the energy-related features collectively form a 7-dimension vector, which includes {*mean, std, median, mad, sum, max, min*}.

- **Self-supervised Learning (SSL) Features**

In the recent past, there has been a surge of research interest in self-supervised learning (SSL). By pre-training a prototype model on huge amounts of unlabeled data, a handful of SSL approaches have shown promising results on diverse low-resource tasks like CAPT, with their focus mainly on MDD [23][24][25]. To the best of our knowledge, there is little prior work on adopting SSL to complement pronunciation assessment [26]. In order to instantiate the multi-view modeling of 3M for pronunciation assessment, we further explore three dominant SSL approaches to examine their effectiveness and feasibility in this work.

*Wav2vec 2.0*: wav2vec 2.0 [27] is a unified and pre-trained model that consists of a feature encoder, a context network, and a quantization module, which are jointly learned in a combination of mask prediction and contrastive learning manner. Furthermore, wav2vec 2.0 achieves state-of-the-art results on phone recognition which is highly related to CAPT. Another study also found that its latent representation renders rich phonetic information [28]. Consequently, we investigate the utility of this model in this work.

*HuBERT*: with the same structure as wav2vec 2.0 but different training strategies, Hidden unit BERT (HuBERT) [29] has been shown to outperform wav2vec 2.0 in various tasks. The main difference between wav2vec 2.0 and HuBERT is that the latter one only applies the prediction loss over the masked region and adopts an iterative refinement strategy on predicting target labels, which uses K-means on either MFCC based acoustic features or the hidden units of Transformer to learn discriminative acoustic representations.

*WavLM*: as an extension of HuBERT, WavLM [30] was designed to learn the masked prediction and speech denoising capabilities simultaneously and obtained significant results on the full-stack downstream speech tasks of speech processing. To be more specific, WavLM adds gated relative position bias to the Transformer structure, and the noisy and overlapped speech are simulated as extra data augmentations in the pre-training phase.

Notably, we extract the above different kinds of SSL features, respectively, by averaging over time frames to align with other phone-level acoustic features, and these features collaborate to provide multiple views on the L2 learner's pronunciation. As for the integration of these multiple views, there are several fusion mechanisms to be considered. In this work, we employ a simple concatenation of all features and subsequently use a projection layer to encapsulate these features. The whole fusion procedure can be formulated as follows:

$$x = [E_{gop}, E_{dur}, E_{eng}, E_{w2v2}, E_{hubert}, E_{wavlm}] \quad (4)$$

$$E_{multi-view} = Dense(x) \quad (5)$$

| Model | Phone Score | | Word Score (PCC) | | | Utterance Score (PCC) | | | | |
|---|---|---|---|---|---|---|---|---|---|---|
| | MSE ↓ | PCC ↑ | Accuracy ↑ | Stress ↑ | Total ↑ | Accuracy ↑ | Completeness ↑ | Fluency ↑ | Prosody ↑ | Total ↑ |
| RF [2] | 0.130 | 0.440 | - | - | - | - | - | - | - | - |
| SVR [2] | 0.160 | 0.450 | - | - | - | - | - | - | - | - |
| Lin et al. [32] | - | - | - | - | - | - | - | - | - | 0.720 |
| LSTM [13] | 0.089 ±0.000 | 0.591 ±0.003 | 0.514 ±0.003 | **0.294** ± **0.012** | 0.531 ±0.004 | 0.720 ±0.002 | 0.076 ±0.086 | 0.745 ±0.002 | 0.747 ±0.005 | 0.741 ±0.002 |
| GOPT [13] | 0.085 ±0.001 | 0.612 ±0.003 | 0.533 ±0.004 | 0.291 ±0.030 | 0.549 ±0.002 | 0.714 ±0.004 | 0.155 ±0.039 | 0.753 ±0.008 | 0.760 ±0.006 | 0.742 ±0.005 |
| *Self-supervised learning approach* | | | | | | | | | | |
| Kim et al. [26] | - | - | - | - | - | - | - | 0.780 | 0.770 | - |
| 3M | **0.078** ±0.001 | **0.656** ±0.005 | **0.598** ±0.005 | 0.289 ±0.033 | **0.617** ±0.005 | **0.760** ±0.004 | **0.325** ±0.141 | **0.828** ±0.006 | **0.827** ±0.008 | **0.796** ±0.004 |

**Table 1.** Comparison of performance with various approaches on speechocean762.

where $E_{gop}$ denotes the GOP features, $E_{dur}$ and $E_{eng}$ denotes the prosodic features of duration and energy, respectively. $E_{w2v2}, E_{hubert}, E_{wavlm}$ represents the SSL features derived from wav2vec 2.0, HuBERT, and WavLM, respectively. $Dense(\cdot)$ is the dense projection layer. This way, the multi-view representation $E_{multi-view}$ will form a 24-dimensionsal embedding for subsequent modeling.

- **Vowel/consonant positional embedding**

In pronunciation assessment, the text prompt for a sentence to pronounce is known in advance. Since different phones have their own characteristics, we can make use of the canonical phones of the text prompt and provide useful information for jointly modeling, i.e., the canonical phone embedding. In addition to the one-hot canonical phone embedding used in GOPT, we develop a vowel/consonant positional embedding (taking inspiration from [12]) as extra phonological feature due to the fact that the stress typically occurs on vowels. By simply using a look-up table of vowels and consonants, we can convert the canonical phone sequence to the corresponding embedding sequences of vowels and consonants. As illustrated in Figure 1, if a canonical phone is vowel, we denote the position with "V"; if it is consonant, then with "C". Subsequently, the canonical phone embedding and proposed vowel/consonant positional embedding are projected to same dimension as the $E_{multi-view}$. We then add these features with the trainable positional embedding and input it to the Transformer encoder.

III. EXPERIMENTS

A. Dataset

To validate the efficacy of our proposed method, we use speechocean762 [2], which is an open-source dataset for pronunciation assessment. The speechocean762 dataset can be downloaded from OpenSLR[1], which contains 5,000 English utterances from 250 non-native speakers, where the speakers range from adults to children. One of the main characteristics of speechocean762 is that the rich annotation information is provided. Specifically, the annotations are scores of pronunciation proficiency in many aspects and at different granularities (viz. utterance-level, word-level, and phone-level annotation labels), and each of them is evaluated manually by five experts independently, all following the same rubrics. For each utterance, there are five utterance-level scores including accuracy, fluency, completeness, prosody, and total scores (ranging from 0-10). For each word, it provides three word-level scores including accuracy, stress, and total scores (ranging from 0-10). For each phone, it offers an accuracy score (ranging from 0-2). On the whole, a total of 9 labels for different granularities and quality-aspect scores are provided. As suggested by [13], we normalize utterance-level and word-level scores to the same scale of the phone score (0-2) for training.

B. Experimental Setup

For a fair comparison, we use the same DNN-HMM acoustic model[2] as [13] to extract 84-dimensional GOP features, which is based on factorized time-delay neural network (TDNN-F) and trained with Librispeech 960-hour data using the widely-used Kaldi recipe. To verify the effectiveness of our proposed multi-view approach, we keep all training hyperparameters the same as the original GOPT including the objective function, and all the experiments are repeated for five trials with different random seeds. The final results are reported with the mean and standard deviation, which are the Pearson Correlation Coefficient (PCC) and mean square error (MSE), respectively. For the SSL features, we use pre-trained wav2vec 2.0[3] [27], HuBERT[4] [29], and WavLM[5] [30] model from HuggingFace [31] to extract the corresponding SSL features, and their dimensions are all set to be 1,024. Because the discrepancy of the dimensionality between SSL and other features (GOP and prosodic features) is quite large, we empirically found that randomly dropout a portion of each SSL features with probability $p_{drop}$ before combination with other features could help the model learn well and prevent it from overfitting. The $p_{drop}$ is set to 0.1 in our experiments.

---
[1] https://www.openslr.org/101
[2] https://kaldi-asr.org/models/m13
[3] https://huggingface.co/facebook/wav2vec2-large-xlsr-53
[4] https://huggingface.co/facebook/hubert-large-ll60k
[5] https://huggingface.co/microsoft/wavlm-large

| Setting | Phone Score | | Word Score (PCC) | | | Utterance Score (PCC) | | | | |
|---|---|---|---|---|---|---|---|---|---|---|
| | MSE ↓ | PCC ↑ | Accuracy ↑ | Stress ↑ | Total ↑ | Accuracy ↑ | Completeness ↑ | Fluency ↑ | Prosody ↑ | Total ↑ |
| *Text Embedding* | | | | | | | | | | |
| Baseline* | 0.085 ±0.001 | 0.612 ±0.003 | 0.533 ±0.004 | 0.291 ±0.030 | 0.549 ±0.002 | 0.714 ±0.004 | 0.155 ±0.039 | 0.753 ±0.008 | 0.760 ±0.006 | 0.742 ±0.005 |
| + Vowel Embed | 0.084 ±0.001 | 0.616 ±0.003 | 0.537 ±0.004 | **0.310** ±**0.034** | 0.553 ±0.007 | 0.717 ±0.004 | 0.110 ±0.149 | 0.758 ±0.004 | 0.756 ±0.003 | 0.745 ±0.004 |
| *Prosodic Features* | | | | | | | | | | |
| + Duration | 0.083 ±0.000 | 0.620 ±0.002 | 0.543 ±0.005 | 0.292 ±0.045 | 0.559 ±0.006 | 0.720 ±0.003 | 0.178 ±0.148 | 0.769 ±0.008 | 0.766 ±0.003 | 0.747 ±0.006 |
| + Energy | 0.082 ±0.001 | 0.626 ±0.004 | 0.549 ±0.004 | 0.292 ±0.025 | 0.565 ±0.005 | 0.723 ±0.005 | 0.220 ±0.124 | 0.779 ±0.007 | 0.778 ±0.004 | 0.749 ±0.006 |
| *Self-supervised Learning Features (add independently)* | | | | | | | | | | |
| + wav2vec2.0 | 0.082 ±0.001 | 0.626 ±0.006 | 0.553 ±0.004 | 0.289 ±0.026 | 0.569 ±0.003 | 0.731 ±0.007 | 0.313 ±0.117 | 0.779 ±0.007 | 0.776 ±0.004 | 0.757 ±0.003 |
| + WavLM | 0.081 ±0.001 | 0.639 ±0.006 | 0.566 ±0.006 | 0.276 ±0.050 | 0.583 ±0.007 | 0.742 ±0.008 | 0.105 ±0.137 | 0.802 ±0.011 | 0.804 ±0.006 | 0.770 ±0.004 |
| + HuBERT | 0.081 ±0.000 | 0.635 ±0.004 | 0.588 ±0.008 | 0.253 ±0.024 | 0.604 ±0.008 | 0.760 ±0.007 | 0.118 ±0.138 | **0.830** ±**0.005** | 0.819 ±0.004 | 0.793 ±0.006 |
| *Multi-view Representation* | | | | | | | | | | |
| 3M | **0.078** ±**0.001** | **0.656** ±**0.005** | **0.598** ±**0.005** | 0.289 ±0.033 | **0.617** ±**0.005** | **0.760** ±**0.004** | **0.325** ±**0.141** | 0.828 ±0.006 | **0.827** ±**0.008** | **0.796** ±**0.004** |

**Table 2.** The ablation study of the proposed method. ∗ denotes the base model GOPT with the canonical phone embedding, and the top-down results are incremental except the self-supervised features experiments.

### C. Overall performance

In the first set of experiments, we discuss the overall performance of our approach in comparison with six well-practiced methods, including Random forest regressor (RF) [2], Support vector regressor (SVR) [2], a transfer learning-based deep feature approach [32], multi-aspect and multi-granularity model (denoted by LSTM) [13], GOPT [13], and an SSL-based method proposed by [26]. Among them, LSTM, GOPT, and 3M adopt the same acoustic model trained on the Librispeech dataset to extract the GOP features. The corresponding results are shown in Table 1, from which we can make several observations. First, in relation to other DNN-enabled approaches, RF and SVR are relatively lightweight regression-based methods, which lead to poor performance. Secondly, although the deep feature method proposed by [32] can obtain some improvements compared to the GOP features in their experiments, the model is only designed to output a total utterance score where the rich annotations of other aspects of pronunciation proficiency are not sufficiently utilized. Contrastively, LSTM and GOPT, which exploit the multi-task learning scheme to learn a model to perform multi-aspect and multi-granularity assessments, deliver superior performance over [32]. Third, by means of the self-attention mechanism, the GOPT model can better learn the contextual information between consecutive phones in an utterance, resulting in better performance than LSTM on most of the assessment measures. Finally, in addition to property of multi-aspect and multi-granularity modeling, the proposed 3M approach extends to include multi-view acoustic, prosodic, and phonological features for pronunciation assessment. By leveraging these features, 3M can obtain significant results on several assessment measures in terms of PCC, particularly for the suprasegmental-level assessment measures such as the fluency and prosody scores, over GOPT and the previous SSL-based method [26]. Rather than using HuBERT as an independent acoustic input [26], the reason that the performance of 3M can go beyond Kim et. al may due to the integration of the multiple-view features, which confirm our hypothesis.

### D. Ablation Study

In the second set of experiments, we turn to examine the factors that will have influence on the performance of our proposed 3M model through ablation studies. As depicted in Table 2, the GOPT model with canonical phone embeddings is set as the baseline model here, and we add one factor at a time to investigate the variation of the performance. Meanwhile, we retain the total scores of each level as high as possible.

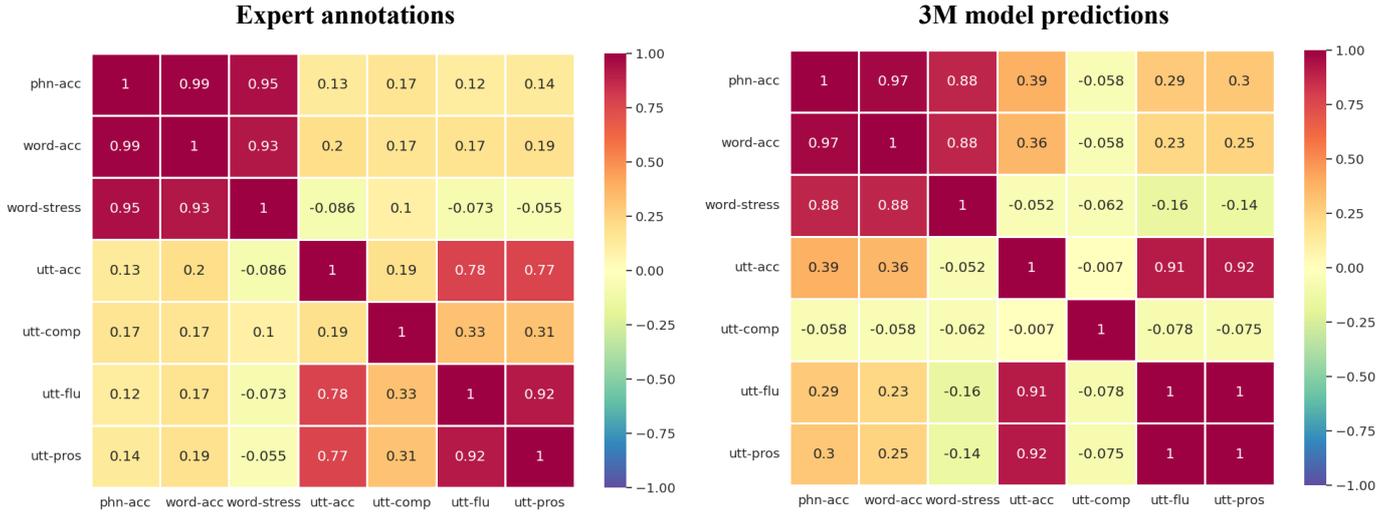

**Fig 2.** Correlation matrices on expert annotations and the 3M model predictions.

In addition to the canonical phone embeddings, we first apply the vowel/consonant positional embeddings as a kind of extra phonological information to augment the baseline model. By applying the positional information of vowels and consonants, we can improve not only the stress score but also the overall scores of all levels in terms of PCC, and it also shows that the position of vowels and consonants are correlated to the stress, which is in line with our intuition.

Furthermore, we incorporate the prosodic features (viz. duration and energy) into the model. Apparently, by taking prosodic features into account, the model has better performance of suprasegmental-level (prosodic-level) assessment on the fluency and prosody scores. Adding these two additional acoustic features has a trade-off that the stress score suffers a slight deterioration, whereas the word- and utterance-level overall scores are improved.

To inspect the utility of leveraging self-supervised learning for the pronunciation assessment task, we further integrate wav2vec 2.0, HuBERT, and WavLM features separately into our model. As shown in Table 2, all three SSL features can increase the overall assessment performance, especially for the total score at the utterance level. Overall, HuBERT is the dominant SSL feature on this task compared to other approaches, and it can offer a considerable improvement on the fluency score. However, the stress score decreases when we introduce additional SSL features (the same as the prosodic features). The poor performance of the stress score and completeness score may result from the extremely imbalanced distribution in the training data. In addition, the standard deviations of these two scores are also much higher than others. This phenomenon is in line with that observed in [13].

Finally, by taking full advantage of above-mentioned acoustic and phonological features, our proposed method can provide multi-view learning to integrate them and achieve the best results on several assessment measures of different granularities ranging from phone to utterance.

*E. Comparative Analysis*

To better understand the correlation between any pair of assessment measures, we conduct a comparative analysis by visualizing the correlation matrix separately on the expert annotations and the 3M model predictions. To this end, we average the phone- and word-level scores of each utterance to align with the utterance-level scores and then calculate the corresponding PCC. The results are presented in Figure 2. For simplicity, we ignore the total scores of each level in this experiment.

Investigating the results in Figure 2, the main findings are as follows. First, the top left and the bottom right areas of the human annotations and the model predictions show a similar tendency, where the phone-level and word-level scores are highly correlated to each other while the utterance-level measures only correlate well to the utterance level on its own. This may indicate that the proposed 3M model can achieve approximate pronunciation assessments as that done by human experts.

Second, there are mismatches between the expert annotations and the model predictions. In the first two rows of the correlation matrix, we find some relations across the phone-, word-, and utterance-level measures in the proposed 3M model. Interestingly, the phone-level accuracy and the word-level accuracy of the expert annotations have low correlations to the utterance-level scores. Based on these observations, we assume the 3M model may learn the correlations between different granularities through the self-attention mechanism during the training phase. However, the PCC values between phone (or word) accuracy and utterance-level scores are less pronounced, and this may imply the incomplete modeling of these granularities. Instead of using the self-attention mechanism on all granularities, a hierarchical modeling structure or a selective attention mechanism may be more suitable.

## IV. CONCLUSION

In this study, we have presented a multi-view, multi-aspect and multi-granularity approach (dubbed 3M), a unified model for automatic pronunciation assessment. Experiments conducted on the speechocean762 benchmark dataset have demonstrated that this novel modeling approach can considerably improve the pronunciation assessment performance at three different levels of granularity (phone-, word- and utterance-level), especially in scoring speaking fluency and speech prosody, which is highly correlated to the proficiency of L2 language learners. Therefore, we believe that this effective approach can provide a promising avenue for future research on CAPT developments. In our future work, we plan to explore the feasibility of equipping 3M with a hierarchical modeling structure for pronunciation assessment.